\documentclass[prb,twocolumn,showpacs]{revtex4}
\usepackage{txfonts}

\begin{document}
\preprint{ACK1021}
\title{Comment on ``Interface tension of Bose-Einstein condensates'' by Bert~Van~Schaeybroeck,
Phys.~Rev.~A \textbf{78}, 023624-9 (2008)}
\author{Todor~M.~Mishonov}\email[E-mail: ]{mishonov@gmail.com}
\affiliation{Department of Theoretical Physics, Faculty of Physics, 
St. Clement of Ohrid University at Sofia,\\ 
5 J. Bourchier Blvd., BG-1164 Sofia, Bulgaria }


\date{April 11, 2015}

\begin{abstract}
The purpose of the comment is to point out that the leading term of the
Ginzburg-Landau nonanalytical correction to the interface tension of 
Bose-Einstein condensates with strong segregation and the surface tension
of extreme type-I superconductors are described by a common coefficient
derived from the universal equation for the phase boundary. 
The agreement between the numerical value of the coefficients 
gives a hint that this can be an exact result which deserves to be checked.
The outcome will be of interest for physicists working in both fields.
\end{abstract}

\maketitle

Recently, the interface tension of the Bose-Einstein condensates attracted significant attention
and the commented article \cite{Schaeybroeck08} is one of the theoretical studies of the problem.
The author states that the main result of his paper\cite{Schaeybroeck08} is the formula for the interface tension
Eq.~(3), which in the used notations can be rewritten as
\begin{eqnarray}
  \gamma_{12}&=&A^*(\xi_1+\xi_2)P-B^*P\nonumber\varsigma_0\\
 &&-\left(\frac{\xi_1}{\xi_2}+\frac{\xi_2}{\xi_1}\right)
 \left[\frac{C^*}{\left(\sqrt{K}\right)^1}
 +\frac{D^*}{\left(\sqrt{K}\right)^2}+\dots\right] P\varsigma_0.
\end{eqnarray}
The first coefficient
$A^*=4\sqrt{2}/3$ coincides with the classical result of Ginzburg-Landau\cite{Ginzburg50} (GL),
The second coefficient 
$B^*=4\times0.514\approx 2.06$
also agrees within the accuracy of the numerical calculations 
with the analytical result\cite{Mishonov88} 
\begin{eqnarray}
\label{UniversalPhaseBoundaryEqn}
&&
\mathrm{d}_{\tau}^2\ \mathcal{X}(\tau)
=\mathcal{X}^2(-\tau)\mathcal{X}(\tau),\quad 
\mathcal{X}(-\infty)=0, \quad 
\left. \mathrm{d}_{\tau} 
\mathcal{X}\right|_{\tau=+\infty}=1,\nonumber\\
&&
 B^*\equiv2^{9/4}\int_{-\infty}^\infty
(1-\mathrm{d}_{\tau}\mathcal{X})\,
\mathrm{d}_{\tau}\mathcal{X}\,
\mathrm{d}\tau, \quad \tau\equiv \hat{z}/2^{1/4}.
\end{eqnarray}
for the superconductor surface tension\cite{Mishonov88}
\begin{eqnarray}
 \gamma=A^*\xi P-\left[
 B^*+ \tilde{C}^*\varkappa
 + \tilde{D}^*\varkappa^2+\dots\,\right]P\sqrt{\xi\lambda},
 \quad \varkappa=\frac{\lambda}{\xi}.
\end{eqnarray}
This numerical coincidence reveals an important new relation between the surface tension of the extreme type-I superconductors and Bose gases with strong segregation which deserves further analysis.

For superconductors, $\xi$ is the temperature dependent coherence length, 
$\lambda$ is the temperature dependent penetration depth,
$P=B_\mathrm{c}^2/2\mu_0$, 
is the magnetic pressure, $B_c$ is the temperature dependent critical magnetic field,
$\mu_0$ is the permeability of the vacuum, 
$\varsigma_0=\sqrt{\xi\lambda},$
and $\varkappa$ is the GL parameter.
We consider that the coefficient $B^*$ is common for both  superfluids 
(the adjacent Boze gases and the type-I superconductors) 
and it is actually the action of an instanton solution 
of the universal GL equations\cite{Ginzburg50}
\begin{eqnarray}
 &&\mathrm{d}_{\tau}^2\ \mathcal{X}
 =\mathcal{Y}^2\mathcal{X},\qquad
 \mathrm{d}_{\tau}^2\ \mathcal{Y}
 =\mathcal{X}^2\mathcal{Y},\\
 && \mathcal{X}(-\infty)=\mathcal{Y}(+\infty)=0,\quad
 \left.\mathrm{d}_{\tau}\mathcal{X}\right|_{\tau=\infty}=1
 =- \left.\mathrm{d}_{\tau}\mathcal{Y}\right|_{\tau=-\infty}.
 \nonumber
\end{eqnarray}
We have to consider $\tau$ as the time for two dimensional motion of a fictitious particle
in the potential $-\frac12\mathcal{X}^2\mathcal{Y}^2$. The corresponding 
mechanical problem is depicted in Fig.~5. of Ref.~[\onlinecite{Mishonov88}]. 
The further coefficients $C^*$, $D^*$, $\tilde{C}^*$, and $\tilde{D}^*$
of the G-L analytical expansions in power of $K^{-1/2}$ or $\varkappa$
can be calculated as a perturbation on the instanton action. 

For the flat geometry of the domain wall the phase invariance of the superfluid order parameter is irrelevant.
The universal constant 
$B^*=2.056347\dots$, 
describes the energy of the domain wall for strongly repulsing order parameters in framework of the original Landau theory\cite{Landau37} for the second order phase transitions. 
For applicability of this result is necessary energy density to have power expansion with respect of order parameters and their first derivatives and the influence of the fluctuations to be small. 
In the spirit of the Landau theory one and the same mathematical problem can be applied for different physical systems. 


\end{document}